\documentclass[fleqn,usenatbib]{mnras}

\usepackage{newtxtext,newtxmath}
\usepackage[T1]{fontenc}

\DeclareRobustCommand{\VAN}[3]{#2}
\let\VANthebibliography\thebibliography
\def\thebibliography{\DeclareRobustCommand{\VAN}[3]{##3}\VANthebibliography}

\usepackage{graphicx}	% Including figure files
\usepackage{amsmath}	% Advanced maths commands

\usepackage{ulem}

\usepackage{multirow}
\usepackage{threeparttable}
\newcommand{\uas}  {$\mu$as}
\newcommand{\uasy}  {$\mu$as~yr$^{-1}$}
\newcommand{\HIP}   {\textsc{Hipparcos}}
\newcommand{\Gaia}  {\textit{Gaia}}

\title[VLBI observing strategies for CRF Link]{VLBI Astrometry of Radio Stars to Link Radio and Optical Celestial Reference Frames: Observing Strategies}

\author[Zhang et al.]{Jingdong Zhang,$^{1,2}$
Bo Zhang,$^{1}$\thanks{E-mail: zb@shao.ac.cn}
Shuangjing Xu,$^{1,3}$
Niu Liu,$^{4}$
Wen Chen,$^{5,6}$
Hao Ding,$^{1}$
Pengfei Jiang,$^{7}$\newauthor
Yan Sun,$^{1}$
Jinqing Wang,$^{1}$
Lang Cui,$^{7}$
Shiming Wen,$^{1}$
Xiaofeng Mai,$^{1,2}$
Jinling Li,$^{1}$
Fengchun Shu,$^{1}$\newauthor
and Yidan Huang$^{1}$
\\
$^{1}$Shanghai Astronomical Observatory, Chinese Academy of Sciences, 80 Nandan Road, Shanghai 200030, People’s Republic of China\\
$^{2}$University of Chinese Academy of Sciences, No.19 (A) Yuquan Rd, Shijingshan, Beijing 100049, People’s Republic of China\\
$^{3}$Korea Astronomy and Space Science Institute, 776 Daedeok-daero,
Yuseong-gu, Daejeon 34055, Republic of Korea\\
$^{4}$School of Astronomy and Space Science, Key Laboratory of Modern Astronomy and Astrophysics (Ministry of Education), Nanjing University, \\Nanjing 210023, People’s Republic of China\\
$^{5}$Yunnan Observatories, Chinese Academy of Sciences, Kunming 650216, Yunnan, People’s Republic of China\\
$^{6}$Yunnan Key Laboratory of the Solar Physics and Space Science, Kunming 650216, People’s Republic of China\\
$^{7}$Xinjiang Astronomical Observatory, Chinese Academy of Sciences, 150 Science 1-Street, Urumqi 830011, People’s Republic of China
}

\date{Accepted 2024 March 6. Received 2024 February 29; in original form 2024 January 11}

\pubyear{2024}

\begin{document}
\label{firstpage}
\pagerange{\pageref{firstpage}--\pageref{lastpage}}
\maketitle

\begin{abstract}The \Gaia\ celestial reference frame (\Gaia-CRF) will benefit from a close assessment with independent methods, such as Very Long Baseline Interferometry (VLBI) measurements of radio stars at bright magnitudes.
However, obtaining full astrometric parameters for each radio star through VLBI measurements demands a significant amount of observation time.
This study proposes an efficient observing strategy that acquires double-epoch VLBI positions to measure the positions and proper motions of radio stars at a reduced cost.
The solution for CRF link compatible with individual VLBI position measurements is introduced, and the optimized observing epoch scheduling is discussed.
Applying this solution to observational data yields results sensitive to sample increase or decrease, yet they remain consistently in line with the literature at the 1-$\sigma$ level.
This suggests the potential for improvement with a larger sample size.
Simulations for adding observations demonstrate the double-epoch strategy reduces CRF link parameter uncertainties by over $30\%$ compared to the five-parameter strategy.
\end{abstract}


%%%%%%%%%%%%%%%%%%%%%%%%%%%%%%%%%%%%%%%%%%%%%%%%%%

%%%%%%%%%%%%%%%%% BODY OF PAPER %%%%%%%%%%%%%%%%%%

\section{Introduction}
\label{sect:intro}

In this era of thriving multi-band astronomy, the establishment and maintenance of a unified celestial reference frame covering all bands is of utmost importance for various research fields such as astronomy, geodesy, deep-space navigation, and more.
Currently, the two main bands in which each has established a high-precision celestial reference frame are radio and optical bands.
At the radio band, the latest realization of the International Celestial Reference Frame \citep[ICRF;][]{1998AJ....116..516M} adopted by the International Astronomical Union (IAU) is ICRF3, based on VLBI measurements of accurate coordinates for 4536 extragalactic radio sources with a coordinate noise floor of 30\,\uas\ \citep{2020A&A...644A.159C}.
At the optical band, the third data release of \Gaia\ mission \citep{2016A&A...595A...1G}, \Gaia\ DR3 \citep{2023A&A...674A...1G}, contains the astrometric parameters for more than 1.6 million quasar-like sources, defining the \Gaia\ Celestial Reference Frame 3 (\Gaia-CRF3) in the optical band with comparable noise floor to ICRF3 \citep{2022A&A...667A.148G}.
The conventional axes of the International Celestial Reference System (ICRS) are defined by ICRF, so \Gaia-CRF needs to be aligned to ICRF for consistency.
It is now aligned with ICRF3 through common extragalactic objects to an uncertainty of about $7\,\mu \mathrm{as}$ \citep{2022A&A...667A.148G}.

Although \Gaia\ aims to provide a globally consistent reference frame for all types of objects, there still exist subtle differences among them, depending on the magnitude, color, position on the celestial sphere, and other factors that produce small shifts of image centroids \citep{2018A&A...616A...2L}.
There is an indication that the bright CRF ($G$ band magnitude $G\lesssim$ 13) of \Gaia\ DR2 rotates with respect to the quasars by more than 100 \uasy~\citep{2018A&A...616A...2L, 2018ApJS..239...31B}.
This deviation is likely caused by a deficiency in the astrometric instrument calibration model \citep[Appendix B]{2020A&A...633A...1L}.
For bright ($G\lesssim$ 13) and faint ($G>$ 13) sources, different ``window classes'' (WCs) of the calibration model are used.
The WCs are schemes for sampling the pixels around a detected source on the charge-coupled device (CCD), and the image centroids derived from different WCs may include systematic errors.
The \Gaia-CRF3 \citep{2022A&A...667A.148G} took this rotation into consideration and applied a spin correction $\omega=(-16.6, -95.0, +28.3)\,\mathrm{\mu as\ yr}^{-1}$ on three axes $[X,Y,Z]$ derived from the proper motion differences between \HIP\ \citep{2007ASSL..350.....V} and \Gaia\ DR2 \citep{2018A&A...616A...1G} during its production \citep{2021A&A...649A...2L}.
The uncertainty in the link of the \HIP\ celestial reference frame (HCRF) to ICRF at epoch J1991.25 was $\pm$0.6\,mas in each axis \citep{1997A&A...323..620K}.
Verifying the bright \Gaia-CRF through the comparison with HCRF is mainly limited by the position error in HCRF, so this method will be of little use in future \Gaia\ data releases \citep{2020A&A...633A...1L}.

Therefore, other inspections, including both internal and external ones, are needed to realize a consistent CRF at different magnitudes.
As an example of internal inspections, \citet{2021A&A...649A.124C} characterized magnitude-dependent systematics in the proper motion of bright sources using resolved binaries and open cluster members spanning above and below $G=13$, and the systematic spin parameters at magnitude bin $12.75<G<13$ are $\omega=(+34.9, +68.9, -2.9)\pm(1.7, 2.1, 1.9)\,\mathrm{\mu as\ yr}^{-1}$.
As for external inspections, which are beneficial for the alignment of optical and radio CRFs, objects observable in both bands are needed.
Since most quasars are faint in the optical band ($G\geq 14$), other kinds of objects are required to estimate the correction parameters of the bright \Gaia-CRF, and VLBI astrometry of radio stars is the most direct way \citep{1997A&A...323..620K,1999A&A...344.1014L}.
To this end, \citet{2020A&A...633A...1L, 2020A&A...637C...5L} linked bright \Gaia-CRF2 to ICRF3 with VLBI astrometric data of the best-fitting 26 radio stars, and obtained a set of link parameters of $\varepsilon=(-19, +1304, +553)\pm(158, 349, 135)\,\mathrm{\mu as},\ \omega=(-68, -51, -14)\pm(52, 45, 66)\,\mathrm{\mu as\ yr}^{-1}$ for orientation and spin parameters, respectively.
After the release of \Gaia-CRF3, because it has improved both the number and precision of sources used to establish the reference frame compared with \Gaia-CRF2, the CRF link parameters at the bright end also need to be re-estimated.
\citet{2022AstL...48..790B} assembled a sample of 84 radio stars with proper motion measurements, and derived a spin of $\omega=(+60, +80, -100)\pm(60, 70, 80)\,\mathrm{\mu as\ yr}^{-1}$ between \Gaia-CRF3 and ICRF3.
\citet{2023A&A...676A..11L} linked \Gaia-CRF3 and ICRF3 using radio stars with improved models of stellar motion and expanded VLBI results, and found that a residual spin of $\omega=(+22, +65, -16)\pm(24, 26, 24)\,\mathrm{\mu as\ yr}^{-1}$ is still significant for bright sources, suggesting that the proper motion correction based on the \HIP\ data involved in \Gaia\ DR3 is not effective enough.

The advantage of using VLBI observations of radio stars to determine the link parameters between the bright Gaia-CRF and ICRF is its independence and ability to estimate the orientation difference at a given epoch.
It is crucial to conduct the VLBI observations before the end of \Gaia\ mission for an accurate orientation parameter estimation.
As a part of our project aims to improve the link accuracy of the bright \Gaia-CRF and ICRF by an order of magnitude, in addition to making use of archive data, we have also observed about a dozen radio stars with VLBI and obtained/updated their astrometric parameters~\citep[e.g.,][]{2023MNRAS.524.5357C}.
However, the number of available radio stars remains limited, and the cost of observation is very high.
For typical VLBI astrometric observations, more than four epochs spanning one year are required to solve such many parameters to obtain five-parameter astrometric data, i.e., Right Ascension (R.A.), Declination (Dec.), parallax, and proper motion (in R.A. and Dec.).
On the other hand, because of their faintness in the radio band, detecting and locating radio stars requires very long integration times.
Therefore, more efficient observing strategies are needed to increase the number of available radio stars.

In this paper, we propose to improve the observing efficiency with a new observing strategy, the double-epoch strategy.
The mathematical formulation of a compatible CRF link solution for this strategy is described in Sect. \ref{sect:solution}.
Then we applied this solution to existing observational data in Sect. \ref{sect:application}.
In Sect. \ref{sect:sim}, we generated a simulated dataset, assessed and compared different observing strategies, and estimated the effect of adding different numbers of new radio stars.
The method to optimally schedule epochs for the double-epoch observation is described in Sect. \ref{sect:epoch}.
Finally, we summarize and look forward in Sect. \ref{sect:summary}.

\section{Solution}
\label{sect:solution}
The double-epoch strategy is to observe each radio star for only two epochs, approximately separated by an integer number of years, so that two individual positions can be obtained.
The purpose of the integer-year time interval is to cancel out parallactic offsets, so an unbiased proper motion measurement can be obtained, which is essential to the spin parameter estimation.
Here, we give a solution on how to incorporate the paired position measurements into CRF link.

The data for a radio star may contain a variety of different forms if new observations (paired position measurements) are added to the historical data.
In this case, the solution for five-parameter input data presented in \citet[Sect.~2]{2020A&A...633A...1L} needs to be slightly modified to be compatible with combined forms of input data, including five-parameter astrometric data and individual positions, etc.
The modified solution is referred to as the ``compatible solution'', and we provide the mathematical formulation of it here in outline.

Suppose there is no translation between the two CRFs, and only a time-dependent rotational offset $\boldsymbol{\varepsilon}(t)$ exists.
Here we assume the angular velocity $\boldsymbol{\omega}$ of the rotation is constant, i.e., $\boldsymbol{\varepsilon}(t)=\boldsymbol{\varepsilon}(T)+\boldsymbol{\omega}\Delta t$, where $\Delta t$ is the time difference between VLBI epoch $t$ and \Gaia\ reference epoch $T$.
Visualizing ICRF and \Gaia-CRF as two sets of orthogonal unit vectors $[\boldsymbol{X},\boldsymbol{Y},\boldsymbol{Z}]$ and $[\tilde{\boldsymbol{X}},\tilde{\boldsymbol{Y}},\tilde{\boldsymbol{Z}}]$ separately, the first-order coordinate differences between the two CRFs can be expressed as
\begin{equation}
\begin{aligned}
    (\alpha-\tilde{\alpha}) \cos \delta &=+\varepsilon_{X} \cos \alpha \sin \delta+\varepsilon_{Y} \sin \alpha \sin \delta-\varepsilon_{Z} \cos \delta\ , \\
    \delta-\tilde{\delta} &=-\varepsilon_{X} \sin \alpha+\varepsilon_{Y} \cos \alpha\ ,
\end{aligned}
\label{eq:coor1}
\end{equation}
where $\boldsymbol{\varepsilon}(T)=[\varepsilon_{X}(T), \varepsilon_{Y}(T), \varepsilon_{Z}(T)]^{\prime}$ denotes the orientation parameters at epoch $T$.
The proper motion differences are given by
\begin{equation}
\label{eq:coor2}
\begin{aligned}
    \mu_{\alpha*}-\tilde{\mu}_{\alpha*} &=+\omega_{X} \cos \alpha \sin \delta+\omega_{Y} \sin \alpha \sin \delta-\omega_{Z} \cos \delta\ , \\
    \mu_{\delta}-\tilde{\mu}_{\delta} &=-\omega_{X} \sin \alpha+\omega_{Y} \cos \alpha\ ,
\end{aligned}
\end{equation}
where $\alpha*$ denotes $\alpha \cos \delta$, $\boldsymbol{\omega} = [\omega_{X}, \omega_{Y}, \omega_{Z}]^{\prime}$ denotes the spin parameters.

The input data consists of two parts, the \Gaia\ part and the VLBI part.
Suppose there are $m$ stars in total, the form of the \Gaia\ data for source $i\ (i=1\cdots m)$ is five-parameter astrometric data $\boldsymbol{g}_i=[\tilde{\alpha} *_i, \tilde{\delta}_i, \varpi_i, \tilde{\mu}_{\alpha *i}, \tilde{\mu}_{\delta i}]^{\prime}$ and corresponding $5\times 5$ covariance matrix $\boldsymbol{C}_i$.
The VLBI part for source $i$ may consist of $n_{i}$ data items $\{\boldsymbol{f}_{ij},\ j=1\cdots n_{i}\}$ which may have different forms, and the dimension of vector $\boldsymbol{f}_{ij}$ is $d_{ij}\times 1$.
For five-parameter form, $\boldsymbol{f}_{ij}=[\alpha *_{ij}, \delta_{ij}, \varpi_{ij}, \mu_{\alpha *{ij}}, \mu_{\delta {ij}}]^{\prime}$ and corresponding covariance matrix $\boldsymbol{V}_{ij}$ is a $5\times 5$ matrix;
For individual position, $\boldsymbol{f}_{ij}=[\alpha *_{ij}, \delta_{ij}]^{\prime}$ and corresponding covariance matrix $\boldsymbol{V}_{ij}$ is a $2\times 2$ matrix.

The parameters to be estimated include the six unknown CRF link parameters $\boldsymbol{x}=[\varepsilon_{X}(T), \varepsilon_{Y}(T), \varepsilon_{Z}(T), \omega_{X}, \omega_{Y}, \omega_{Z}]^{\prime}$ and the corrections to \Gaia\ five-parameter astrometric data
\begin{equation}
   \boldsymbol{y}_i=
   \begin{bmatrix}
    \begin{aligned}
      &\Delta{\alpha*}_i \\
      &\Delta\delta_i \\
      &\Delta\varpi_i \\
      &\Delta\mu_{\alpha*i} \\
      &\Delta\mu_{\delta i}
    \end{aligned}
   \end{bmatrix} \ .
\end{equation}
These parameters can be estimated using a least-square algorithm.
The computation procedure of the algorithm is briefly described here.
The loss function to minimize is
\begin{multline}
    \label{eq:loss}
   Q(\boldsymbol{x},\{\boldsymbol{y}_{i}\})=\sum_{i=1\cdots m} \{(\boldsymbol{y}_{i}-\boldsymbol{K}_{i} \boldsymbol{x})^{\prime} \boldsymbol{C}_{i}^{-1}(\boldsymbol{y}_{i}-\boldsymbol{K}_{i} \boldsymbol{x}) \\
   +\sum_{j=1\cdots n_{i}}[(\boldsymbol{\Delta} \boldsymbol{f}_{ij}- \boldsymbol{M}_{ij}\boldsymbol{y}_{i})^{\prime} \boldsymbol{V}_{ij}^{-1}(\boldsymbol{\Delta} \boldsymbol{f}_{ij}- \boldsymbol{M}_{ij}\boldsymbol{y}_{i})] \} \ ,
\end{multline}
where
\begin{equation}
    \boldsymbol{K}_{i}=\begin{bmatrix}
        \mathrm{c} \alpha_{i} \mathrm{s} \delta_{i} & \mathrm{s} \alpha_{i} \mathrm{s} \delta_{i} & -\mathrm{c} \delta_{i} & 0 & 0 & 0 \\
        -\mathrm{s} \alpha_{i} & \mathrm{c} \alpha_{i} & 0 & 0 & 0 & 0\\
        0 & 0 & 0 & 0 & 0 & 0\\
        0 & 0 & 0 & \mathrm{c} \alpha_{i} \mathrm{s} \delta_{i} & \mathrm{s} \alpha_{i} \mathrm{s} \delta_{i} & -\mathrm{c} \delta_{i}\\
        0 & 0 & 0 & -\mathrm{s} \alpha_{i} & \mathrm{c} \alpha_{i} & 0
        \end{bmatrix}
\end{equation}
is the coefficient matrix from Eqs. \eqref{eq:coor1}-\eqref{eq:coor2}, $\boldsymbol{M}_{ij}$ is the Jacobian matrix of the propagation model of \Gaia\ data at $\boldsymbol{y}_{i}=0$, and $\boldsymbol{\Delta f}_{ij}=\boldsymbol{f}_{ij}-\boldsymbol{M}_{ij}\boldsymbol{g}_{i}$ is the difference between the observed data and propagated \Gaia\ parameters.

For different forms of VLBI data, different $\boldsymbol{M}_{ij}$, $\boldsymbol{\Delta} \boldsymbol{f}_{ij}$ and $\boldsymbol{V}_{ij}$ are used.
In the case of five-parameter data,
\begin{equation}
    \boldsymbol{M}_{ij}=\begin{bmatrix}
    1 & 0 & 0 & \Delta t_{ij} & 0 \\
    0 & 1 & 0 & 0 & \Delta t_{ij} \\
    0 & 0 & 1 & 0 & 0 \\
    0 & 0 & 0 & 1 & 0 \\
    0 & 0 & 0 & 0 & 1
    \end{bmatrix} \ ,
\end{equation}
$\boldsymbol{\Delta} \boldsymbol{f}_{ij}$ is a $5\times 1$ vector, and $\boldsymbol{V}_{ij}$ is a $5\times 5$ matrix.

In the case of individual position, the propagation model is based on linear proper motion and parallactic ellipse calculated from ephemeris.
The Cartesian coordinates $[X_{\odot ij}, Y_{\odot ij}, Z_{\odot ij}]$ (unit: AU) of the Sun relative to the Earth at each epoch are provided by the Jet Propulsion Laboratory's (JPL) DE440 ephemeris \citep{2021AJ....161..105P}, based on which we have
\begin{equation}
    \label{eq:prop_start}
    \boldsymbol{M}_{ij}=\begin{bmatrix}
    1 & 0 & Y_{\odot ij} \mathrm{c} \tilde{\alpha}-X_{\odot ij} \mathrm{s} \tilde{\alpha} & \Delta t_{ij} & 0\\
    0 & 1 & Z_{\odot ij} \mathrm{c} \tilde{\delta}-X_{\odot ij} \mathrm{c} \tilde{\alpha} \mathrm{s} \tilde{\delta}-Y_{\odot ij} \mathrm{s} \tilde{\alpha} \mathrm{s} \tilde{\delta} & 0 & \Delta t_{ij}
    \end{bmatrix} \ ,
\end{equation}
$\boldsymbol{\Delta} \boldsymbol{f}_{ij}$ is a $2\times 1$ vector, and $\boldsymbol{V}_{ij}$ is a $2\times 2$ matrix.

Radial velocity, Roemer delay, and other effects are currently not taken into consideration in either case of propagation models, because of their relatively small scale compared to the present measurement accuracy.
Other forms of VLBI data are also possible to be included in the solution by modifying the corresponding $\boldsymbol{M}_{ij}$, $\boldsymbol{\Delta} \boldsymbol{f}_{ij}$ and $\boldsymbol{V}_{ij}$.

Minimizing $Q$ can be achieved by solving the normal equation
\begin{equation}
\label{eq:x_solve}
   \left(\sum_{i=1...m} \boldsymbol{K}_{i}^{\prime} \boldsymbol{C}_{i}^{-1} \boldsymbol{K}_{i}\right) \boldsymbol{x}-\sum_{i=1...m} \boldsymbol{K}_{i}^{\prime} \boldsymbol{C}_{i}^{-1} \boldsymbol{y}_{i}=0 \ ,
\end{equation}
\begin{multline}
\label{eq:y_solve}
   [\boldsymbol{C}_{i}^{-1}+\sum_{j=1\cdots n_{i}}( \boldsymbol{M}_{ij}'\boldsymbol{V}_{ij}^{-1}\boldsymbol{M}_{ij} )] \boldsymbol{y}_{i} = \\
   \boldsymbol{C}_{i}^{-1} \boldsymbol{K}_{i} \boldsymbol{x} + \sum_{j=1\cdots n_{i}}(\boldsymbol{M}_{ij}'\boldsymbol{V}_{ij}^{-1} \boldsymbol{\Delta f}_{ij}) \ , \ i=1\cdots m.
\end{multline}
Substitute Eq. \eqref{eq:y_solve} into Eq. \eqref{eq:x_solve}, $\hat{\boldsymbol{x}}$ can be first solved from
\begin{equation}
   \label{eq:reduced_norm}
   \left(\sum_{i=1...m} \boldsymbol{N}_{i}\right) \boldsymbol{x}=\sum_{i=1...m} \boldsymbol{b}_{i} \ ,
\end{equation}
where
\begin{equation}
   \begin{aligned}
      \boldsymbol{N}_{i} &= \boldsymbol{K}_{i}^{\prime} \boldsymbol{C}_{i}^{-1}(\boldsymbol{K}_{i} - \boldsymbol{H}_{i}^{-1} \boldsymbol{C}_{i}^{-1} \boldsymbol{K}_{i}) \ , \\
      \boldsymbol{b}_{i} &= \boldsymbol{K}_{i}^{\prime} \boldsymbol{C}_{i}^{-1} \boldsymbol{H}_{i}^{-1} \cdot \sum_{j=1\cdots n_{i}}(\boldsymbol{M}_{ij}^{\prime} \boldsymbol{V}_{ij}^{-1} \boldsymbol{\Delta f}_{ij}) \ ,
   \end{aligned}
\end{equation}
and
\begin{equation}
   \boldsymbol{H}_{i} = \boldsymbol{C}_{i}^{-1} + \sum_{j=1\cdots n_{i}} (\boldsymbol{M}_{ij}^{\prime}\boldsymbol{V}_{ij}^{-1}\boldsymbol{M}_{ij}) \ .
\end{equation}
Then $\hat{\boldsymbol{y}}_{i}$ can be solved from Eq.\eqref{eq:y_solve}.
The covariance matrix of $\hat{\boldsymbol{x}}$ is $(\sum_{i=1...m} N_{i})^{-1}$, and the loss can be calculated from
\begin{equation}
    \begin{aligned}
        Q(\boldsymbol{x}) &= \sum_{i=1...m} Q_{i}(\boldsymbol{x}) \ , \\
        Q_{i}(\boldsymbol{x}) &= \sum_{i=1...n_{i}} Q_{ij}(\boldsymbol{x}) \ , \\
        Q_{ij}(\boldsymbol{x}) &= (\boldsymbol{\Delta f}_{ij} - \boldsymbol{M}_{ij}\boldsymbol{K}_{i}\boldsymbol{x})^{\prime}\boldsymbol{D}_{ij}^{-1}(\boldsymbol{\Delta f}_{ij} - \boldsymbol{M}_{ij}\boldsymbol{K}_{i}\boldsymbol{x}) \ ,
    \end{aligned}
\end{equation}
where
\begin{equation}
    \begin{aligned}
        \boldsymbol{D}_{ij} = \boldsymbol{V}_{ij}+\boldsymbol{M}_{ij}\boldsymbol{C}_{i}\boldsymbol{M}_{ij}^{\prime} \ .
    \end{aligned}
\end{equation}
So the total loss, loss of each star, and the loss of each data item of the star can all be calculated.
To assess the goodness of fit, the reduced chi-square can be calculated as
\begin{equation}
    \label{eq:chi2}
    \begin{aligned}
        \chi^2_{\mathrm{red}} &= \frac{Q(\boldsymbol{x})}{\nu} \ , \\
        \nu &= \sum_{i=1\cdots m}\nu_i \ , \\
        \nu_i &=\sum_{j=1\cdots n_{i}}d_{ij} \ .
    \end{aligned}
\end{equation}

% E and Omega definition
The amount of information on $\boldsymbol{\varepsilon}(T)$ and $\boldsymbol{\omega}$ contributed by each source can be derived from the first and last three diagonal elements of $\boldsymbol{N}_{i}$ respectively:
\begin{equation}
    E_i=\underset{\varepsilon}{\mathrm{trace}}(\boldsymbol{N}_{i}) \quad \mathrm{and} \quad \Omega_i=\underset{\omega}{\mathrm{trace}}(\boldsymbol{N}_{i}) \ .
\end{equation}
For more detail about the interpretation of the mathematics, see \citet{2020A&A...633A...1L}.

\section{Application to observational data}
\label{sect:application}
In this section, we applied the compatible solution to observational data as a basis for the following simulation.
41 radio stars are included in the final solution in \citet[see Table 1]{2020A&A...633A...1L}.
\citet[see Table E.1, E.3]{2023A&A...676A..11L} added one-epoch measurements for 32 radio stars to the sample, and applied correlations to the positions of 29 stars in \citet{2020A&A...633A...1L} for homogenization, for example, update their reference source positions to ICRF3 coordinates.
55 radio stars in total are included in the solution of \citet{2023A&A...676A..11L}, and we also use this dataset and adopt their correlations here.
The uncertainties of the 32 one-epoch measurements we adopted are the inflated uncertainties they provided, taking into account both thermal and systematic errors.
We also make use of new measurements of two radio stars, HD 199178 and AR Lac, given in \citet{2023MNRAS.524.5357C}.
The optical counterparts are collected from \Gaia\ DR3, and we adopted the parallax zero-point correction recipe provided by \citet{2021A&A...649A...4L}, which takes $G$-band magnitude, color, and ecliptic latitude into account.

\begin{figure*}
    \includegraphics[width=2\columnwidth]{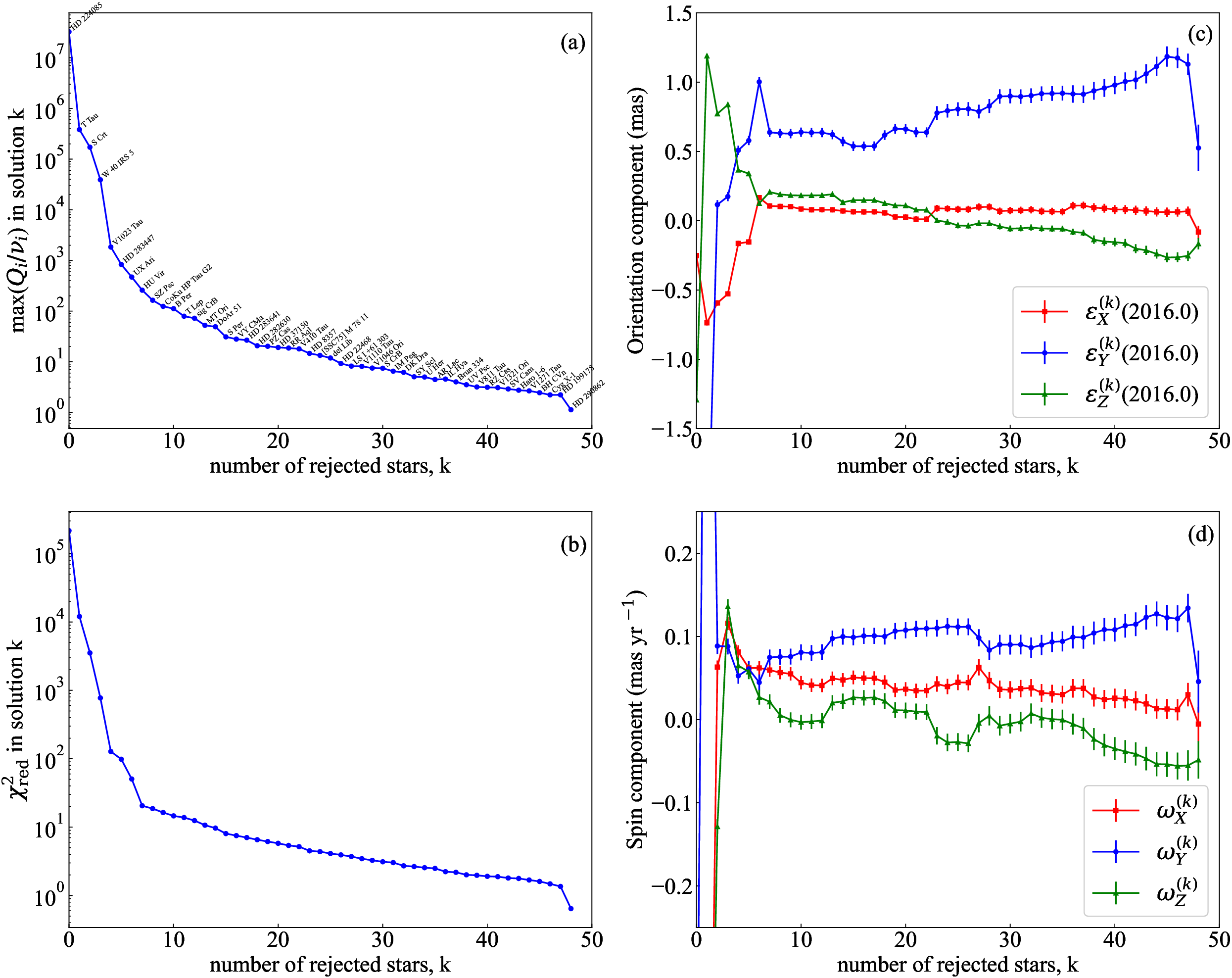}
    \caption{The evolution of loss and CRF link parameters with rejected star number $k$ grows on the full dataset.
    (a) The max $Q_{i}(\boldsymbol{x})$ in the solution.
    (b) $\chi^2_{\mathrm{red}}$ of the solution.
    (c) Estimated orientation parameters.
    (d) Estimated spin parameters.
    Error bars are formal uncertainties directly derived from the covariance matrix $(\sum_{i=1...m} N_{i})^{-1}$.}
    \label{fig:evolution}
\end{figure*}

The direct application of the CRF link solution to the dataset will be seriously biased by outliers.
To eliminate radio stars with high loss, we applied a similar iteration procedure as \citet{2020A&A...633A...1L, 2023A&A...676A..11L} used.
The solution is first applied to the entire dataset, then the star with the highest $Q_{i}/\nu_i$ is discarded and the solution is applied again to the remaining stars.
The iteration is repeated for $k=0, 1, \cdots$ rejected stars, and the evolution of $\mathrm{max}(Q_{i}/\nu_i)$, $\chi^2_{\mathrm{red}}$ and $\boldsymbol{x}$ is shown in Figure \ref{fig:evolution}.
In Panel (a), the downward trend of max $Q_{i}/\nu_i$ flattens out after $k=10$ (in logarithmic coordinates, same below), while in Panel (b), a similar turning point occurs at $k=6$.
The estimated CRF link parameters shown in Panels (c) and (d) have non-negligible changes as $k$ increases, especially the spin parameters.
We arbitrarily choose $k=16$ here.

Three subsets of data are used to evaluate the effect of adding new observations to the radio star samples collected by \citet{2020A&A...633A...1L}: 41 stars from \citet{2020A&A...633A...1L} only; 55 stars from both \citet{2020A&A...633A...1L} and \citet{2023A&A...676A..11L}; 55 stars from the entire dataset.
The same iteration procedure as above is also applied to the first two subsets, and $k=11$ and $16$ are chosen for them respectively.
The results are shown in Table \ref{table:apply-to-obs}.
\begin{table*}
    \centering
    \caption{Results of the application to \Gaia\ DR3 and VLBI observational data}
    \label{table:apply-to-obs}
    \begin{threeparttable}
        \begin{tabular}{cccccccccc}
            \hline
            Subset & $m$ & $k$ & \multicolumn{3}{c}{Orientation at $T={\rm J}2016.0$ (mas)} & \multicolumn{3}{c}{Spin (mas yr$^{-1}$)} & $\chi^2_{\mathrm{red}}$ \\
            & & & $\varepsilon_X(T)$ & $\varepsilon_Y(T)$ & $\varepsilon_Z(T)$ & $\omega_X$ & $\omega_Y$ & $\omega_Z$ & \\
            \hline
            1 & 41 & 11 & $-0.050\pm 0.023$ & $+0.230\pm 0.047$ & $+0.090\pm 0.017$ & $+0.010\pm 0.009$ & $+0.060\pm 0.011$ & $-0.007\pm 0.011$ & $8.256$ \\
            2 & 55 & 16 & $-0.032\pm 0.021$ & $+0.257\pm 0.043$ & $+0.100\pm 0.016$ & $+0.019\pm 0.009$ & $+0.063\pm 0.010$ & $+0.004\pm 0.010$ & $7.104$ \\
            3 & 55 & 16 & $+0.064\pm 0.019$ & $+0.538\pm 0.034$ & $+0.148\pm 0.015$ & $+0.050\pm 0.008$ & $+0.101\pm 0.010$ & $+0.026\pm 0.009$ & $7.513$ \\
            \hline
            Adopted & 55 & 16 & $+0.064\pm 0.053$ & $+0.538\pm 0.092$ & $+0.148\pm 0.041$ & $+0.050\pm 0.023$ & $+0.101\pm 0.026$ & $+0.026\pm 0.025$ & \\
            \hline
        \end{tabular}
        \begin{tablenotes}    
            \footnotesize               
            \item[~] The three subsets are:\\
            (1) 41 stars from \citet{2020A&A...633A...1L}, with position corrections from \citet{2023A&A...676A..11L};\\
            (2) 55 stars from both \citet{2020A&A...633A...1L} and \citet{2023A&A...676A..11L};\\
            (3) Same 55 stars as (2) but with new measurements for HD 199178 and AR Lac from \citet{2023MNRAS.524.5357C}.\\
            Formal uncertainties of the parameters are calculated from the inverse normal matrix.
            The ``Adopted'' line is the result of subset (3) but with uncertainties normalized by $\chi^2_{\mathrm{red}}$, so that the uncertainties can be compared with the simulation results in Table \ref{table:simu}.
        \end{tablenotes}
    \end{threeparttable}
\end{table*}

The application to the entire dataset gives a result basically consistent with \citet{2021A&A...649A.124C} and \citet{2023A&A...676A..11L}:
Systematic rotation is most significant on the Y-axis, followed by the X-axis, but not on the Z-axis.
However, the consistency is not very significant (merely around the 1-$\sigma$ level).
During the procedure of addition (from subset (1) to (2) and (3)) and removal (the iteration procedure shown in Figure \ref{fig:evolution}) of stars, the CRF link parameters show significant fluctuations.
This can be explained by the small sample size, which is still far from giving a stable CRF link.

The estimated parameters and their uncertainties do not change much with 32 one-epoch measurements added for subset (2) relative to subset (1).
As shown in Table \ref{table:statistics}, the one-epoch observations added in subset (2) only slightly increase the weight of HD 199178 and AR Lac, while the new observations of the two stars given in \citet{2023MNRAS.524.5357C} contribute quite a lot in the solution.
We suggest that multi-epoch measurements perform better in the CRF link, and this will be verified in Sect. \ref{sect:evaluation}.
\begin{table}
    \centering
    \caption{Solution statistics for HD 199178 and AR Lac}
    \label{table:statistics}
    \begin{threeparttable}
        \begin{tabular}{cccccc}
            \hline
            Name & Subset & $E_i$ & $\Omega_i$ & $\nu_i$ & $Q_{i}/\nu_i$ \\
             & & $\mathrm{mas}^{-2}$ & $\mathrm{mas}^{-2}\mathrm{yr}^{2}$ & & \\
            \hline
            \multirow{3}{*}{HD 199178} & 1 & 8.2 & 4062.6 & 5 & 1.05 \\
             & 2 & 17.8 & 4100.6 & 7 & 1.05 \\
             & 3 & 421.8 & 6713.9 & 12 & 5.99 \\
            \hline
            \multirow{3}{*}{AR Lac} & 1 & 6.2 & 3408.3 & 5 & 1.79 \\
             & 2 & 39.6 & 3453.7 & 9 & 2.25 \\
             & 3 & 303.2 & 4177.8 & 14 & 4.21 \\
            \hline
        \end{tabular}
        \begin{tablenotes}    
            \footnotesize               
            \item[~] $E_i$ and $\Omega_i$ are the weights for the determination of orientation and spin parameters, respectively.
            $\nu_i$ is the sum of the dimensions of all data of a star included in the solution.
        \end{tablenotes}
    \end{threeparttable}
\end{table}

\section{Simulation for future observing strategy design}
\label{sect:sim}

The unstable CRF link result shown in \ref{sect:application} indicates the need for a larger sample of available radio stars.
However, as mentioned in Sect. \ref{sect:intro}, to obtain five-parameter astrometric data is high-cost.
Therefore, in this section, we conducted a simulation to compare the three observing strategies: single-epoch, five-parameter, and double-epoch strategies.
For the single-epoch strategy, the positions of stars are only measured once.
For the five-parameter strategy, each star is simulated to be observed over six epochs within a whole year, and the astrometric parameters are estimated from the six individual positions.
For the double-epoch strategy, the positions of stars are measured over two distinct epochs separated by a full year.
To ensure comparable observation costs across the three strategies, we kept the total number of epochs the same.

\subsection{Simulated dataset}
\label{sect:dataset}
The evaluation of the strategies needs a dataset containing three different forms of simulated data:

(1) optical five-parameter astrometric data $\{\boldsymbol{r}_i,\ i=1\cdots m\},\ \boldsymbol{r}_i = [ \alpha*_i, \delta_i, \varpi_i, \mu_{\alpha*_i}, \mu_{\delta_i} ]'$;

(2) individual VLBI positions $\{\boldsymbol{v}_{ij},\ i=1\cdots m,\ j=1\cdots n_{i}\},\ \boldsymbol{v}_{ij} = [ \alpha*_{ij}, \delta_{ij}]'$;

(3) VLBI five-parameter astrometric data $\{\boldsymbol{s}_i,\ i=1\cdots m\},\ \boldsymbol{s}_i = [ \alpha*_i, \delta_i, \varpi_i, \mu_{\alpha*_i}, \mu_{\delta_i} ]'$.

\begin{figure*}
    \includegraphics[width=2\columnwidth]{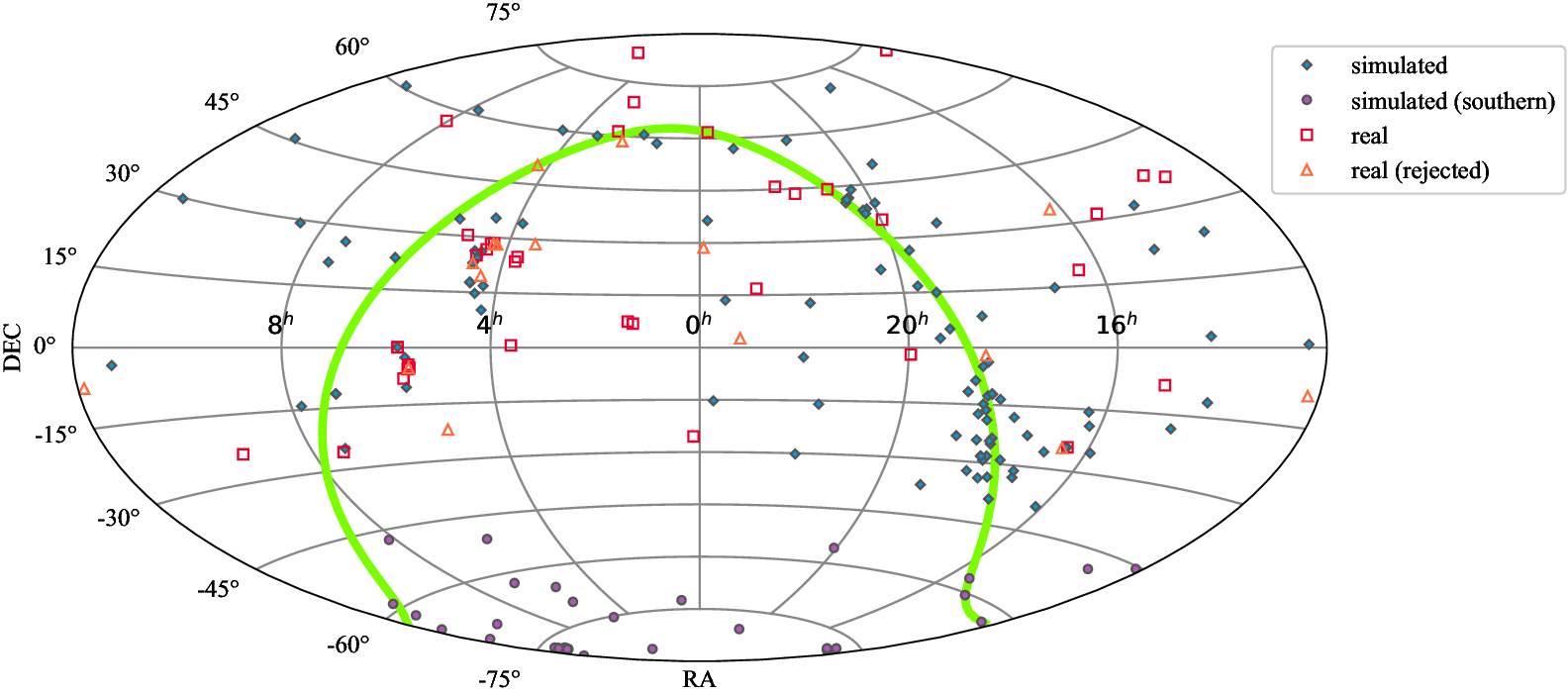}
    \caption{The sky distribution (Aitoff projection in equatorial coordinates) of the radio stars used in this work.
    The blue solid diamonds denote the 111 radio stars collected from VLASS, while the purple dots denote the 35 radio stars distributed in the deep-southern sky collected from \citet{2015yCat.8099....0W}.
    The red hollow squares and the orange hollow triangles indicate the 39 accepted and the 16 rejected ``real'' radio stars, respectively.
    The Galactic Equator is plotted in green.}
    \label{fig:star_distri}
\end{figure*}

$m$ is the number of radio stars, while $n$ is the epoch number of each radio star.
Since the goal of the simulation is to help design future observations, the generation of simulated dataset requires a list of radio stars that are observable with VLBI but not yet observed.
Here we first collected radio stars (those exhibiting non-thermal radio emission detectable by VLBI, such as RS Canum Venaticorum variables and X-ray binaries) from the Very Large Array Sky Survey (VLASS, \citealt{2021ApJS..255...30G}) and \Gaia.
Specifically, we crossmatched the radio star samples from VLASS with \Gaia\ DR3, and used the \Gaia\ DR3 five-parameter astrometric data at epoch $T_{{\rm Gaia}} = 2016.0$ as the ``true'' data for dataset generating.
However, VLASS only covers the sky area with $\delta>-40^{\circ}$, so we also collected some radio stars distributed in the deep-southern sky from \citet{2015yCat.8099....0W}.
The stars already used in Sect. \ref{sect:application} were excluded.
To evaluate the impact of sky distribution, the 111 radio stars collected from VLASS were classified into dataset A, while the 35 deep-southern radio stars collected from \citet{2015yCat.8099....0W} were classified into dataset B.
In total, $m=146$ radio stars are collected.
Note that the candidate selection criteria only require that the radio flux density and compactness of the radio star are sufficient for VLBI detection and that there is a \Gaia\ counterpart.
So the dataset only reflects the radio star distribution on the celestial sphere, which is shown in Figure \ref{fig:star_distri}, and cannot be directly used as a list of candidate sources for future observations.

The simulation would be meaningless if no noises were added to the data, since all the input data are perfectly matched in CRF connection. 
Although the errors of astrometric parameters include both systematic errors and random errors, in this work, only random errors are considered, because systematic errors have different origins and characteristics, making them difficult to describe with simple models in the simulation.
With Gaussian noise (using the uncertainties given by \Gaia\ DR3 as standard deviation $\sigma$) added to the ``true'' data, the optical five-parameter astrometric data $\{\boldsymbol{r}_i\}$ are generated.
The individual VLBI positions $\{\boldsymbol{v}_{ij}\}$ are generated using the same propagation model as Eq. \eqref{eq:prop_start} shows.
The number $n_{\mathrm{epoch}}=6$ and time distribution of epochs (evenly distributed between 2023.5 and 2024.5) of all stars are kept the same to preserve the inner consistency of the dataset.
Gaussian noise is also added to the VLBI positions and the standard error $\sigma_{\mathrm{VLBI}}=50\,\mathrm{ \mu as}$ of the noise is used as position uncertainty.
This accuracy has been verified to be achievable through MultiView \citep{2017AJ....153..105R} observation at C-band, e.g., \citet{2022ApJ...932...52H,2023ApJ...953...21H}.
VLBI five-parameter astrometric data $\{\boldsymbol{s}_i\}$ were then generated.
Here a least-square algorithm is used to estimate the five astrometric parameters and the corresponding covariance matrices from individual position sequences.

All the data generated above are under the same CRF, so we need to rotate the optical data to another CRF.
The adopted CRF link parameters $\boldsymbol{x}=[\varepsilon_{X}(T_{{\rm \Gaia}}), \varepsilon_{Y}(T_{{\rm \Gaia}}), \varepsilon_{Z}(T_{{\rm \Gaia}}), \omega_{X}, \omega_{Y}, \omega_{Z}]^{\prime}$ shown in Table \ref{table:apply-to-obs} are used in the rotation, and the optical five-parameter data $\{\tilde{\boldsymbol{r}}_i,\ i=1...m\},\ \tilde{\boldsymbol{r}}_i = [ \tilde{\alpha}*_i, \tilde{\delta}_i, \varpi_i, \tilde{\mu}_{\alpha*_i}, \tilde{\mu}_{\delta_i} ]'$ can be calculated as
\begin{multline}
   \tilde{\boldsymbol{r}}_i = \\
   -\begin{bmatrix}
      \mathrm{s} \alpha_{i} \mathrm{s} \delta_{i} & \mathrm{s} \alpha_{i} \mathrm{s} \delta_{i} & -\mathrm{s} \delta_{i} & 0 & 0 & 0 \\
      -\mathrm{s} \alpha_{i} & \mathrm{s} \alpha_{i} & 0 & 0 & 0 & 0 \\
      0 & 0 & 0 & 0 & 0 & 0 \\
      0 & 0 & 0 & \mathrm{s} \alpha_{i} \mathrm{s} \delta_{i} & \mathrm{s} \alpha_{i} \mathrm{s} \delta_{i} & -\mathrm{s} \delta_{i} \\
      0 & 0 & 0 & -\mathrm{s} \alpha_{i} & \mathrm{s} \alpha_{i} & 0
   \end{bmatrix}
   \boldsymbol{x} + \boldsymbol{r}_i \ .
\end{multline}
Now the rotation is complete and the simulated dataset containing $\{\tilde{\boldsymbol{r}}_i\}$, $\{\boldsymbol{v}_{ij}\}$, $\{\boldsymbol{s}_i\}$ is ready for observing strategy evaluation.

\subsection{Strategy evaluation}
\label{sect:evaluation}
The evaluation procedures for the three strategies are discussed separately below:

(1) For the single-epoch strategy, stars are randomly sampled from $\{\tilde{\boldsymbol{r}}_i\}$ and $\{\boldsymbol{v}_{ij}\}$.
Each star has one VLBI position measurement at either $2023.5$ or $2024.5$, and corresponding optical five-parameter astrometric data.

(2) For the five-parameter strategy, stars are randomly sampled from $\{\tilde{\boldsymbol{r}}_i\}$ and $\{\boldsymbol{s}_i\}$.

(3) For the double-epoch strategy, stars are randomly sampled from $\{\tilde{\boldsymbol{r}}_i\}$ and $\{\boldsymbol{v}_{ij}\}$.
VLBI position measurements at both $2023.5$ and $2024.5$ are included.

To assess the effect of adding new observation results to existing data, the simulated samples are combined with the ``real'' data, and the CRF link parameters are then estimated with both five-parameter and double-epoch strategies.

1,000 rounds of sampling were conducted for each strategy, and the mean values of the results are shown in Table \ref{table:simu}.
The first three lines of Table \ref{table:simu} are at the same observation cost so a comparison can be made between the three strategies.
The single-epoch strategy performs significantly worse than the other two strategies, suggesting that multi-epoch measurement is necessary for the CRF link.
Between the multi-epoch strategies, the double-epoch strategy is a more efficient choice, i.e., it obtains lower parameter uncertainties at the same cost, which are over $30\%$ lower than that of the five-parameter strategy.
\begin{table*}
    \centering
    \caption{Results of the simulation evaluating the observing strategies}
    \label{table:simu}
    \begin{threeparttable}
        \begin{tabular}{cccccccccccc}
            \hline
            Strategy & dataset & $n_{\mathrm{star}}$\tnote{*} & $n_{\mathrm{epoch}}$\tnote{*} & $k_{\Gaia}$ & $\chi^2_{\mathrm{red}}$ & \multicolumn{3}{c}{$\sigma_{\varepsilon}$ ($\mathrm{\mu as}$)} & \multicolumn{3}{c}{$\sigma_{\omega}$ ($\mathrm{\mu as\,yr}^{-1}$)} \\
            & & & & & & $X$ & $Y$ & $Z$ & $X$ & $Y$ & $Z$ \\
            \hline
            single-epoch & A+B & 120 & 1 & 1 & 1.080 & 352.9 & 458.5 & 486.8 & 44.5 & 57.7 & 61.7 \\
            five-parameter & A+B & 20 & 6 & 1 & 1.032 & 121.1 & 156.3 & 149.2 & 16.5 & 21.3 & 20.7 \\
            double-epoch & A+B & 60 & 2 & 1 & 1.080 & 82.6 & 101.7 & 96.3 & 11.0 & 13.6 & 13.1 \\
            double-epoch & A & 60 & 2 & 1 & 1.084 & 82.8 & 117.2 & 86.6 & 11.0 & 15.5 & 11.7 \\
            five-parameter & A+B+R & 20 & 6 & 1 & 1.021 & 43.5 & 74.9 & 35.2 & 7.9 & 11.8 & 9.1 \\
            double-epoch & A+B+R & 60 & 2 & 1 & 1.040 & 40.2 & 65.1 & 33.5 & 6.0 & 9.2 & 6.3 \\
            double-epoch & A+B+R & 60 & 2 & 0.5 & 1.035 & 41.1 & 64.0 & 33.8 & 5.4 & 8.3 & 4.9 \\
            \hline
        \end{tabular}
        \begin{tablenotes}    
            \footnotesize               
            \item[~] In the column ``dataset'', A denotes that only radio stars with $\delta>-40^{\circ}$ collected from VLASS are used, A+B denotes that radio stars collected from \citet[$\delta<-40^{\circ}$]{2015yCat.8099....0W} are added, and A+B+R denotes that radio stars are collected from dataset A+B in addition to ``real'' data, thereby evaluating the effect of adding new observations.
            $k_{\Gaia}$ denotes the multiplier for the uncertainties of \Gaia\ astrometric parameters.
            $\chi^2_{\mathrm{red}}$ is the reduced chi-square of the least-square estimation.
            For the first four lines of the table, the total observation costs $n_{\mathrm{star}}\times n_{\mathrm{epoch}}$ are kept the same to compare the efficiency of the strategies.
            Formal uncertainties of the parameters are calculated from the inverse normal matrix.
            Since the uncertainties of the simulated data are consistent with the standard deviation of the Gaussian noise added to the parameters, normalization is not needed for the simulated data, while the covariance matrices of all ``real'' data are scaled by their reduced chi-square to normalize the result.
            \item[*] $n_{\mathrm{star}}$ and $n_{\mathrm{epoch}}$ are the numbers of stars and epochs sampled from the simulated dataset respectively, do not include the ``real'' data used in the last two lines of this table.
        \end{tablenotes}
    \end{threeparttable}
\end{table*}

The comparison between the datasets with and without the deep-southern radio stars shows that the improvement of the uniformity of sky distribution of radio stars helps to reduce the difference in the uncertainties of the CRF link parameters on the three axes.
So it is valuable to add new observations of radio stars distributed in the deep-southern sky.

\begin{figure}
    \includegraphics[width=\columnwidth]{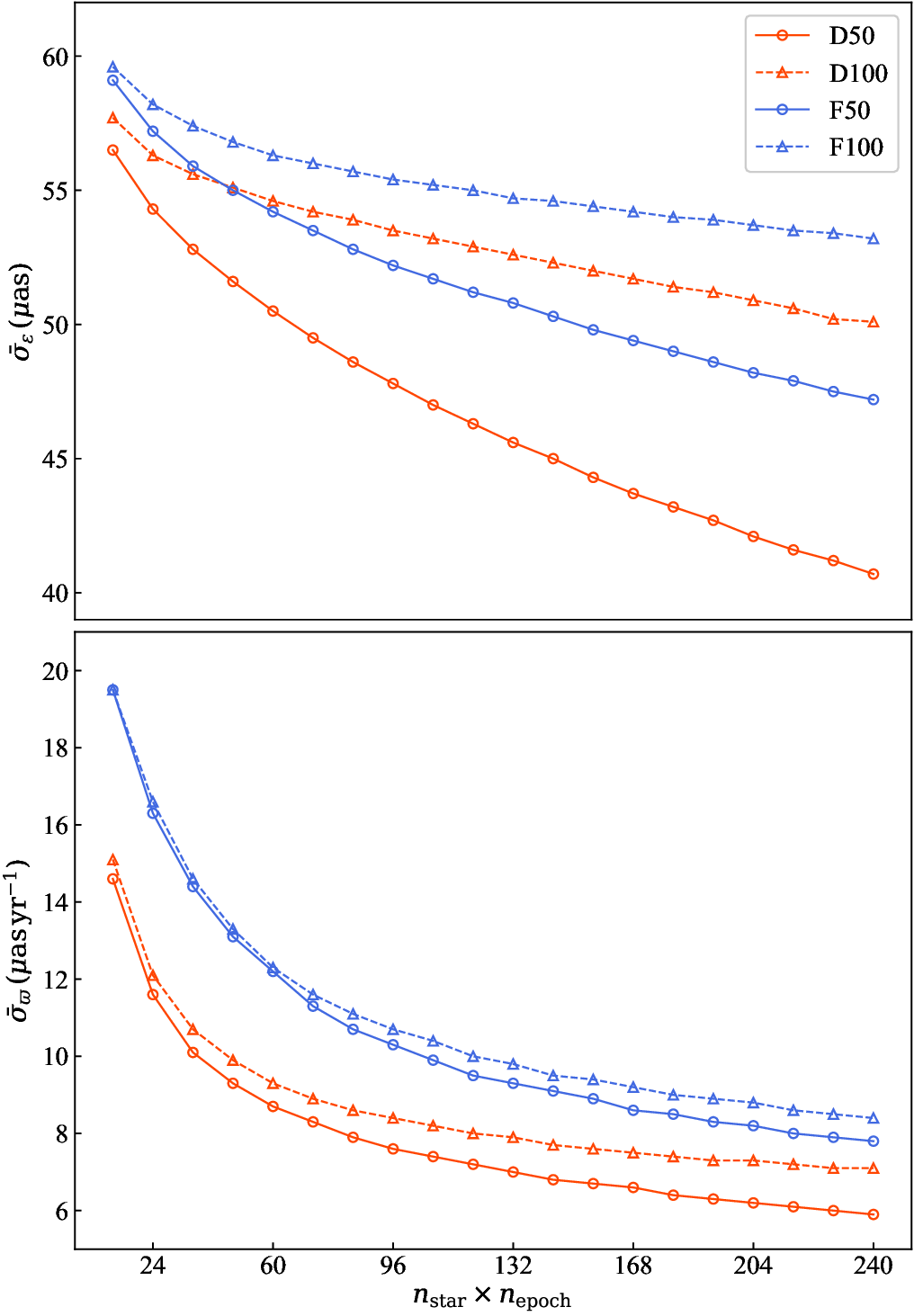}
    \caption{The results of adding different numbers of new stars with the five-parameter and double-epoch strategy to the ``real'' data.
    $n_{\mathrm{star}}\times n_{\mathrm{epoch}}$ is the total epoch number: for the five-parameter strategy, $n_{\mathrm{star}}=6$; for the double-epoch strategy, $n_{\mathrm{star}}=2$.
    $\bar{\sigma}_{\varepsilon}$ (top panel) and $\bar{\sigma}_{\varpi}$ (bottom panel) are the mean uncertainties on three axes of orientation and spin parameters, respectively.
    D50, D100, F50, F100 denote double-epoch strategy with $\sigma_{\mathrm{VLBI}}=50\,\mathrm{\mu as}$, double-epoch strategy with $\sigma_{\mathrm{VLBI}}=100\,\mathrm{\mu as}$, five-parameter strategy with $\sigma_{\mathrm{VLBI}}=50\,\mathrm{\mu as}$, and five-parameter strategy with $\sigma_{\mathrm{VLBI}}=100\,\mathrm{\mu as}$, respectively.
    }
    \label{fig:std}
\end{figure}

We compared the results of adding different numbers of new stars to the ``real'' data with both five-parameter and double-epoch strategies under $\sigma_{\mathrm{VLBI}}=50\,\mathrm{\mu as}$ and $\sigma_{\mathrm{VLBI}}=100\,\mathrm{\mu as}$ respectively, which is shown in Figure \ref{fig:std}.
The results show that the double-epoch strategy is always a better choice as the added epoch number grows, and the orientation parameters are more sensitive to VLBI position uncertainty relative to the spin parameters.
Adding 60 new stars with the double-epoch strategy to the ``real'' data under $\sigma_{\mathrm{VLBI}}=50\,\mu\mathrm{as}$ obtains $25\%$ and $71\%$ lower uncertainties in orientation and spin parameters respectively, while adding 120 new stars further improves these to $34\%$ and $76\%$.

The future \Gaia\ data release will improve astrometric parameter accuracy by 40\% on the precision of all optical data products and a factor of almost three for the proper motions \citep{2021A&A...649A...1G}.
So we tried to multiply all \Gaia\ DR3 parameter uncertainties by a factor $k_{\Gaia}=0.5$ to see how much this will contribute to CRF link accuracy improvement.
The result shows that the contribution is insignificant, indicating that the quantity and quality of VLBI data are more important limiting factors.

Although the simulation is not a perfect reflection of the real situation, the conclusions will help to design future observations of radio stars for the link of the bright \Gaia-CRF to ICRF and quantitatively estimate the effect of adding new observations.

\subsection{Discussion on radio-optical offsets}

The radio-optical offsets caused by the physical properties of radio stars are not considered in the simulation,  such as differences in radiation regions arising from varying radiation mechanisms.
Assuming that the directions and magnitudes of the offsets of different stars are random, a larger sample size will statistically result in averaging out the offsets to zero in the aggregate.

The double-epoch strategy will add about three times more new stars than the five-parameter strategy, therefore in practice the influence of unmodeled offsets will decrease, and the CRF link result will be more robust.
In practice, a large proportion of radio stars are discarded due to their excessive residuals, which is inevitable.
The larger sample size will also allow some of the samples to be discarded, rather than gambling on a few.

Observations before the end of \Gaia\ mission will contribute more to the derivation of the orientation parameters, thus eliminating the instrument error and allowing the true radio-optical offsets to be derived, and the physical nature of radio-optical offsets can then be studied.
Although the observation of \Gaia\ mission is coming to its end, the closer VLBI epoch is to \Gaia\ reference epoch, the larger contribution it will make.
This is another valuable aspect of these observations.

\section{Practical observing epoch scheduling}
\label{sect:epoch}
In this part, we discuss how to practically schedule the observing epochs for the double-epoch strategy.
The strategy is designed not to be affected by parallax, however, the time interval between the two epochs is practically difficult to be exactly an integer number of years, which will bring parallactic displacement.
In the compatible CRF link solution, this displacement is corrected by an offset calculated from \Gaia\ parallax $\varpi_{Gaia}$.
Suppose there exists a bias $\Delta\varpi$ between $\varpi_{Gaia}$ and the ``true'' parallax $\varpi_{\mathrm{true}}$, the displacement would not be perfectly corrected.
This will affect the independence of the CRF link, and reduce the accuracy of the position and proper motion of the radio star.
It is necessary to quantitatively estimate the impact of $\Delta\varpi$, and we give the mathematical form for the estimation here.

The parallactic displacement is caused by the change in the position of the Sun relative to the Earth on an annual cycle.
Since the error estimation here does not need high accuracy, the Earth's orbit can be approximated to be a circle.
So the ecliptic longitude of the Sun at epoch $T$ can be expressed as
\begin{equation}
    \lambda_{\odot}=(\lambda_0+360^{\circ}\mathrm{yr}^{-1}\times T)\bmod 360^{\circ} \ ,
\end{equation}
where $\lambda_0=280^{\circ}$ is an approximation of the ecliptic longitude of the Sun at the beginning of a year.
The Cartesian coordinate $[X, Y, Z]$ (unit: AU) of the Sun can be calculated as
\begin{equation}
\label{eq:sunXYZ}
    \begin{aligned}
    X(T)&=\cos{\lambda_{\odot}} \ , \\
    Y(T)&=\sin{\lambda_{\odot}}\cos{i} \ , \\
    Z(T)&=\sin{\lambda_{\odot}}\sin{i} \ ,
   \end{aligned}
\end{equation}
where $i=23.4^{\circ}$ is an approximation of the obliquity of the ecliptic.
The impact of $\Delta\varpi$ on position can then be derived:
\begin{equation}
\label{eq:pi_pos_1}
      {\Delta \alpha *}_{\varpi}(T)=\Delta\varpi(\sin{\lambda_{\odot}}\cos{i}\cos{\alpha}-\cos{\lambda_{\odot}}\sin{\alpha}) \ ,
\end{equation}
\begin{multline}
\label{eq:pi_pos_2}
      \Delta \delta_{\varpi}(T)=\Delta\varpi(\sin{\lambda_{\odot}}\sin{i}\cos{\delta} \\
      -\cos{\lambda_{\odot}}\cos{\alpha}\sin{\delta}-\sin{\lambda_{\odot}}\cos{i}\sin{\alpha}\sin{\delta}) \ .
\end{multline}
If the time interval $\Delta T$ between the two epochs is close to an integer number of years, a linear approximation can be used to approximately estimate the impact of $\Delta\varpi$ on proper motion.
Differentiate Eqs. \eqref{eq:pi_pos_1}-\eqref{eq:pi_pos_2} by $T$:
\begin{equation}
\label{eq:pi_pm_1}
      {\Delta \alpha *}_{\varpi}(T)'=2\pi\Delta\varpi(\cos{\lambda_{\odot}}\cos{i}\cos{\alpha}+\sin{\lambda_{\odot}}\sin{\alpha}) \ ,
\end{equation}
\begin{multline}
\label{eq:pi_pm_2}
      \Delta \delta_{\varpi}(T)'=2\pi\Delta\varpi(\cos{\lambda_{\odot}}\sin{i}\cos{\delta} \\
      +\sin{\lambda_{\odot}}\cos{\alpha}\sin{\delta}-\cos{\lambda_{\odot}}\cos{i}\sin{\alpha}\sin{\delta}) \ ,
\end{multline}
the impact of $\Delta\varpi$ on proper motion can be expressed as
\begin{equation}
    \begin{aligned}
      {\Delta \mu}_{\alpha * \varpi}(T)'&={\Delta \alpha *}_{\varpi}(T)'\frac{\Delta t}{\Delta T} \ , \\
      {\Delta \mu}_{\delta \varpi}(T)'&=\Delta \delta_{\varpi}(T)'\frac{\Delta t}{\Delta T} \ ,
   \end{aligned}
\end{equation}
% $\Delta t=\lvert \Delta T - \lfloor \Delta T +0.5 \rfloor \rvert$
where $\Delta t=\lvert \Delta T - \mathrm{round} (\Delta T)\rvert$ is the time difference between $\Delta T$ and its proximal integer number of years.
Assuming $\Delta t\le 0.05$, which is feasible in actual epoch scheduling, the $\mu\mathrm{as}\,\mathrm{yr}^{-1}$-scale bias is negligible with current measurement precision ($\sigma_{\varpi}=$ 0.02–0.03 mas at $G<15$, \citet{2021A&A...649A...1G}).

\begin{figure}
    \includegraphics[width=\columnwidth]{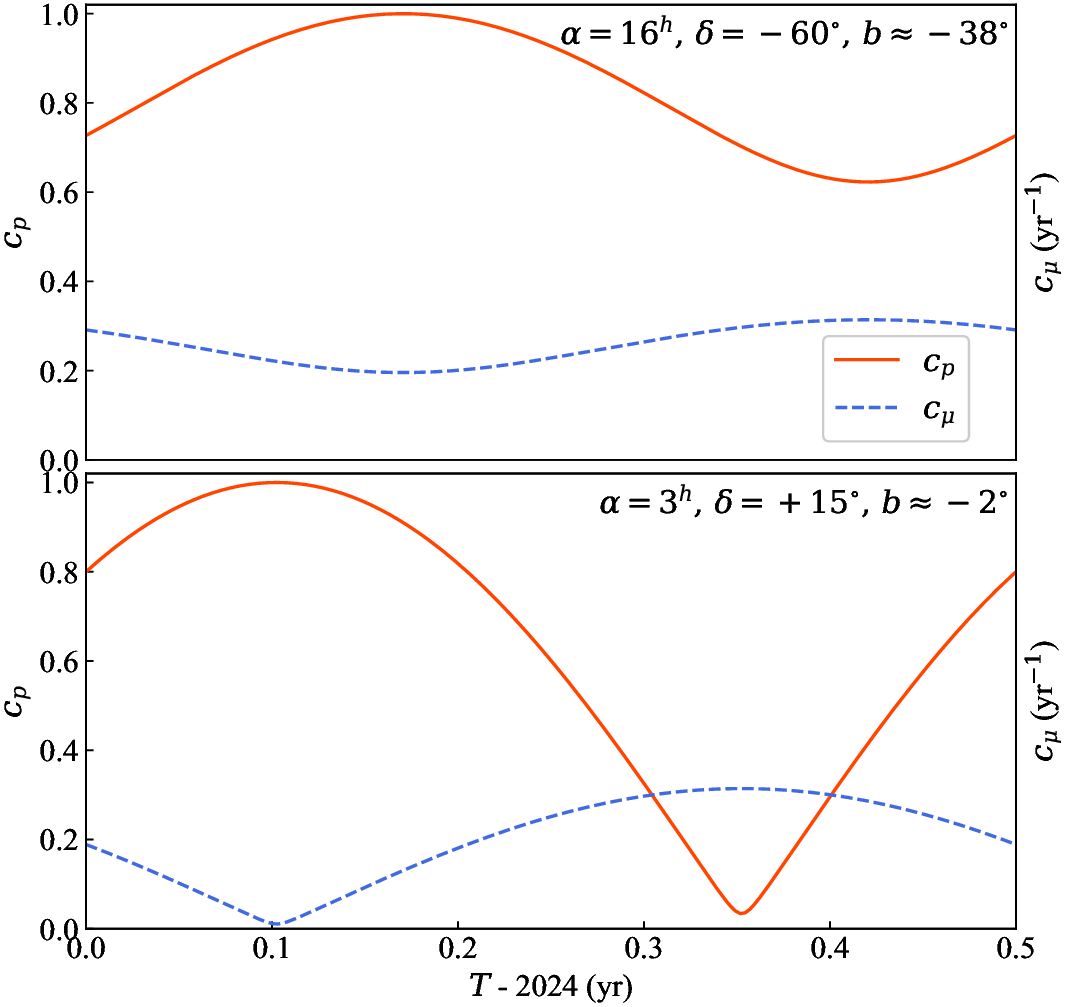}
    \caption{The impact of $\Delta\varpi$ on position $c_p$ and proper motion $c_{\mu}$ as function of epoch $T$.
    Due to the symmetry of the parallactic ellipse, an epoch range of only half a year is shown in the figure.
    $\Delta t/\Delta T$ is fixed to 0.05.
    Top panel: far from the Ecliptic, $\alpha=16^{h}, \delta=-60^{\circ}$, and Ecliptic latitude $b\approx-38^{\circ}$;
    Bottom panel: close to the Ecliptic, $\alpha=3^{h}, \delta=+15^{\circ}$, and $b\approx-2^{\circ}$.
    The red solid line and blue dashed line denote $c_p$ and $c_{\mu}$ respectively.
    }
    \label{fig:coef}
\end{figure}

For epoch $T$ within possible observing time, plotting the trends of the impact of $\Delta\varpi$ will help to find the best epoch.
Two coefficients can be used to quantitatively express the impact of $\Delta\varpi$ on position and proper motion respectively:
\begin{equation}
    \begin{aligned}
      c_{p}&=\Delta\varpi^{-1}\sqrt{{\Delta \alpha *}_{\varpi}(T)^2+\Delta \delta_{\varpi}(T)^2} \ , \\
      c_{\mu}&=\Delta\varpi^{-1}\sqrt{{\Delta \mu}_{\alpha * \varpi}(T)'^{2}+{\Delta \mu}_{\delta \varpi}(T)'^2} \\
      &=2\pi\frac{\Delta t}{\Delta T}\Delta\varpi^{-1}\sqrt{{\Delta \alpha *}_{\varpi}(T)'^{2}+\Delta \delta_{\varpi}(T)'^2}\ .
   \end{aligned}
\end{equation}
$\Delta\varpi^{-1}$ in the equation is for cancelling out $\Delta\varpi$ in Eqs. \eqref{eq:pi_pos_1}-\eqref{eq:pi_pm_2}, therefore $c_{p}$ is a function of epoch $T$ and position $(\alpha,\,\delta)$, while $c_{\mu}$ is in addition related to $\Delta t/\Delta T$.
Two examples are shown in Figure \ref{fig:coef}, with fixed $\Delta t/\Delta T=0.05$ and different positions in the sky. For the one far from the Ecliptic, the ranges of variation of $c_p$ and $c_{\mu}$ are small, and $c_p$ is always higher than $c_{\mu}$ because of the extra coefficient $2\pi\Delta t/\Delta T$ of $c_{\mu}$; while for the other one close to the Ecliptic, $c_p$ and $c_{\mu}$ change significantly with time, and $c_p$ is sometimes lower than $c_{\mu}$.

Therefore, for most cases that stars are far from the Ecliptic, the impact of $\Delta\varpi$ is not very sensitive to epoch $T$;
While for the stars close to the Ecliptic, it is important to carefully choose the epoch $T$ to minimize the impact on position or proper motion depending on the needs:
A smaller $c_p$ will enable the star to contribute more to the orientation parameters, while conversely a smaller $c_{\mu}$ will enable the star to contribute more to the spin parameters.

Since $c_{\mu}$ is also (even more) sensitive to $\Delta t/\Delta T$ other than the choice of epoch $T$, it is possible and better to reduce the impact of $\Delta\varpi$ on proper motion through reducing $\Delta t/\Delta T$: extend the time interval between two epochs $\Delta T$ to more years, and/or limit the time interval even closer to an integer number of years.

As a conclusion, for the minimization of the impact of $\Delta\varpi$, it is advised to choose a pair of epochs with a lower $c_p$, and extend the time interval between epochs and tightening constraints on observation dates to reduce $c_{\mu}$.

\section{Summary and future outlook}
\label{sect:summary}

In this study, we proposed a double-epoch observing strategy for radio star VLBI observation, aiming to validate the consistency of \Gaia-CRF at the bright end with high efficiency.
Firstly, we introduced a CRF link solution that is compatible with various forms of input data, and then applied the solution to currently available data, which result indicates the need for a larger sample size.
We conducted a simulation to compare three strategies for additional observations: single-epoch, five-parameter, and double-epoch strategy, and the results show that the double-epoch strategy is the most efficient one among them.
Specifically, the parameter uncertainties of the double-epoch strategy are over $30\%$ lower than those of the five-parameter strategy at the same observation cost.

We also compared the datasets with and without the radio stars distributed in the deep-southern sky ($\delta<-40^{\circ}$), and the result shows that the improvement of the uniformity of sky distribution of radio stars makes the uncertainties of CRF link parameters on the three axes more even.
We found that the bottleneck of the link accuracy is the quantity and quality of VLBI astrometric data, and the future improvement of \Gaia\ data will not contribute much.
Therefore, the addition of high-quality VLBI measurements is the key to improving the link accuracy.
Under the conditions set in our simulation (epochs close to 2024.0, position accuracy of $\sigma_{\mathrm{VLBI}}=50\,\mu\mathrm{as}$, etc.), if the double-epoch strategy is followed to observe 60 new radio stars, the uncertainties of the orientation and spin parameters will be reduced by $25\%$ and $71\%$, respectively.

We discussed how to schedule new observations based on the double-epoch strategy, and gave a recipe to estimate the impact of possible \Gaia\ parallax bias on position and proper motion measurement.
We found that the priority of epoch choice should be given to reducing the impact on position in most cases, while reducing the impact on proper motion can be more effectively achieved by extending epoch interval and tightening constraints on observation dates.

The new observation can also be combined with archival data, e.g., a radio star was once observed but no proper motion was derived, or the proper motion derived previously is not accurate enough, adding one or two epochs is an effective way to obtain an accurate proper motion measurement.
The archival data with a long time interval between its epoch and the \Gaia\ reference epoch contributes very little to the estimation of orientation parameters, in this case, adding epochs will greatly increase its contribution.
It is also possible to add additional epochs after the double-epoch observation is complete to obtain a full five-parameter astrometric measurement.

Significant improvements in the accuracy of the link of the bright \Gaia-CRF to ICRF would require a substantial number of new observations.
We have proposed observing some radio stars, including both previously observed and new targets, to improve the proper motion accuracy of archive data and increase the number of available radio stars.
In addition, the new facilities anticipated in the near future, such as the Next Generation Very Large Array (ngVLA) and the Square Kilometer Array (SKA), will greatly augment our capacity for observing radio stars by enabling the detection of currently unobservable faint targets.
With a large number of new observations of radio stars, a more precise CRF link model can be adopted, i.e., taking into account effects such as Galactic aberration, magnitude/color dependence of the CRF link parameters, etc.
The double-epoch strategy can also be applied to other frame links, such as the link between the reference frames realized by VLBI astrometry and pulsar timing respectively.

\section*{Acknowledgements}

We would like to thank the referee for very helpful comments and suggestions.
This work is supported by the National Natural Science Foundation of China (NSFC) under grant Nos. U2031212 and 11903079.
This work is also supported by the Strategic Priority Research Program of the Chinese Academy of Sciences, Grant No. XDA0350205.
N. Liu is supported by the National Natural Science Foundation of China (NSFC) under grant Nos. 12103026 and 12373074.
L. Cui is supported by the National SKA Program of China (No. 2022SKA0120102) and the CAS ``Light of West China" Program (No. 2021-XBQNXZ-005).

This work has made use of data from the European Space Agency (ESA) mission
{\it Gaia} (\url{https://www.cosmos.esa.int/gaia}), processed by the {\it Gaia}
Data Processing and Analysis Consortium (DPAC,
\url{https://www.cosmos.esa.int/web/gaia/dpac/consortium}). Funding for the DPAC
has been provided by national institutions, in particular the institutions
participating in the {\it Gaia} Multilateral Agreement.

The \Gaia\ services (\url{https://gaia.ari.uni-heidelberg.de/index.html}) provided by the Astronomisches Rechen-Institut (ARI) of the University of Heidelberg are used in \Gaia\ data retrieval.
The Python package for \Gaia\ DR3 parallax zero-point correction developed by P. Ramos can be found at \url{https://gitlab.com/icc-ub/public/gaiadr3_zeropoint}.
Astropy \citep{astropy2013,astropy2018,astropy2022} and TOPCAT \citep{2005ASPC..347...29T} are used in data analysis.
This work has also made use of the SIMBAD database, operated at CDS, Strasbourg, France \citep[\url{https://simbad.u-strasbg.fr/simbad/}]{simbad}.

%%%%%%%%%%%%%%%%%%%%%%%%%%%%%%%%%%%%%%%%%%%%%%%%%%
\section*{Data Availability}
% The inclusion of a Data Availability Statement is a requirement for articles published in MNRAS. Data Availability Statements provide a standardised format for readers to understand the availability of data underlying the research results described in the article. The statement may refer to original data generated in the course of the study or to third-party data analysed in the article. The statement should describe and provide means of access, where possible, by linking to the data or providing the required accession numbers for the relevant databases or DOIs.

% This work makes use of \Gaia\ data: \url{https://gea.esac.esa.int/archive/}.

% The \Gaia\ data used in this work can be found at: \url{https://gea.esac.esa.int/archive/}.
% The VLBI data used in this work are collected from \citet{2020A&A...633A...1L,2023A&A...676A..11L,2023MNRAS.524.5357C}.

No new data were generated or analysed in support of this research.



\bibliographystyle{mnras}
\bibliography{main} % if your bibtex file is called example.bib

%\begin{thebibliography}{99}
%\end{thebibliography}

%%%%%%%%%%%%%%%%%%%%%%%%%%%%%%%%%%%%%%%%%%%%%%%%%%

%%%%%%%%%%%%%%%%% APPENDICES %%%%%%%%%%%%%%%%%%%%%

% \appendix

% \section{Some extra material}

% If you want to present additional material which would interrupt the flow of the main paper,
% it can be placed in an Appendix which appears after the list of references.

%%%%%%%%%%%%%%%%%%%%%%%%%%%%%%%%%%%%%%%%%%%%%%%%%%

% Don't change these lines
\bsp	% typesetting comment
\label{lastpage}
\end{document}